# Assessing the performance of 8 AI chatbots in bibliographic reference retrieval: Grok and DeepSeek outperform ChatGPT, but none are fully accurate

Álvaro Cabezas-Clavijo & Pavel Sidorenko-Bautista

alvaro.cabezas@unir.net; pavel.sidorenkobautista@unir.net

**Universidad Internacional de La Rioja (UNIR)**

**Avenida de la Paz, 137, Logroño 26006, Spain**

*This manuscript is a preprint and has not undergone peer review. Comments, suggestions, and constructive feedback are welcome and appreciated to help improve the quality and clarity of the final version.*

**ABSTRACT**

This study analyzes the performance of eight generative artificial intelligence chatbots - ChatGPT, Claude, Copilot, DeepSeek, Gemini, Grok, Le Chat, and Perplexity- in their free versions, in the task of generating academic bibliographic references within the university context. A total of 400 references were evaluated across the five major areas of knowledge (Health, Engineering, Experimental Sciences, Social Sciences, and Humanities), based on a standardized prompt. Each reference was assessed according to five key components (authorship, year, title, source, and location), along with document type, publication age, and error count. The results show that only 26.5% of the references were fully correct, 33.8% partially correct, and 39.8% were either erroneous or entirely fabricated. Grok and DeepSeek stood out as the only chatbots that did not generate false references, while Copilot, Perplexity, and Claude exhibited the highest hallucination rates. Furthermore, the chatbots showed a greater tendency to generate book references over journal articles, although the latter had a significantly higher fabrication rate. A high degree of overlap was also detected among the sources provided by several models, particularly between DeepSeek, Grok, Gemini, and ChatGPT. These findings reveal structural limitations in current AI models, highlight the risks of uncritical use by students, and underscore the need to strengthen information and critical literacy regarding the use of AI tools in higher education.

**Introduction**

The effective use of bibliographic references in academic assignments or coursework is a core competency for university students. The ability to identify relevant information sources and apply them correctly in academic writing is essential for developing critical thinking and fostering skills in structuring and communicating complex reasoning. Traditionally, students were required to use academic databases or scholarly search engines to locate appropriate sources. However, since late 2022, the emergence of generative artificial intelligence (GenAI) tools -such as ChatGPT, Gemini, or Copilot- has introduced a new pathway for students to gather such information directly through conversational chatbots.

ChatGPT and related technologies rely on machine learning and natural language processing techniques to generate human-like responses based on text or voice prompts. Their applications range from answering simple queries and retrieving information to producing written or multimedia content (Lund & Wang, 2023; Mishra & Awasthi, 2023). Specifically, within scientific and academic contexts, these tools are increasingly used for streamlining editorial workflows, searching and curating information, and even creating technical, artistic, or intellectual content. This evolution raises important questions about their implications for creativity and originality in student work, the quality of research outcomes, and the and the risk that they may lead to a homogenization in how ideas are expressed and structured in academic contexts (Gutiérrez-Caneda et al., 2023).

In academic writing, relevant bibliographic references are used to support the author's ideas and argumentation. Therefore, students are expected to rely on specialized information sources that provide appropriate documents aligned with their subject of study. However, many students find it challenging to retrieve references through academic databases such as Web of Science, Scopus, or PubMed (Öğrenci, 2013). Although generative models like ChatGPT do not fully meet the complex informational needs of academic contexts (Hersh, 2024), chatbots allow students to query an artificial intelligence (AI) agent directly, bypassing the often cumbersome access protocols of subscription-based databases or the specific syntaxes required by specialized platforms, for which students frequently lack formal training.

In the specific domain of bibliographic reference generation, several studies (Day, 2023; Giray, 2024; Orduña-Malea & Cabezas-Clavijo, 2023; Walters & Wilder, 2023) have shown that chatbots tend to hallucinate -that is, to fabricate references that appear plausible but are in fact false-. This poses a significant problem for users who lack deep familiarity with the bibliographic landscape in their field of study, as is often the case with university students, or for those who accept such references without verifying their existence or appropriateness.

Given that chatbots are now widespread and commonly used within the academic community (Fundación CyD, 2025; Stöhr et al., 2024), this study aims to assess the accuracy of bibliographic references generated by eight general-purpose AI chatbots in their free-access versions, across the five major areas of knowledge. The objective is not only to measure the degree of factual correctness, but also to analyze other relevant aspects for student use, such as the document type of the references provided, their recency, and the intellectual relevance of the cited works. Furthermore, the study seeks to identify whether there is overlap among the references suggested by the different applications.

The novelty of this study lies in offering the first comparative analysis of eight AI chatbots specifically focused on their ability to generate bibliographic references for student academic work. This systematic evaluation makes it possible to identify which models perform better, while also contributing to a deeper understanding of the risks associated with the uncritical use of these tools in academic settings. The study thus aims to provide empirical evidence that may inform both the design of information literacy strategies and the development of educational policies regarding the responsible use of artificial intelligence in higher education.

**Literature Review**

*Generative Artificial Intelligence in Higher Education*

Generative artificial intelligence (GenAI), which has emerged abruptly and rapidly, is characterized by its ability to create new data based on learned patterns, spanning domains such as text and multimedia content (Lund et al., 2023). This capacity relies on deep learning and neural networks, which enable the simulation of human output with increasingly realistic results (Ray, 2023). ChatGPT, arguably the most representative and paradigmatic example of GenAI tools, employs machine learning and natural language processing techniques to generate human-like responses to text-based prompts (Javaid et al., 2023). This is the same underlying technology used by other systems such as Gemini, Claude, or Grok, among others.

The range of applications of these tools includes answering questions of all kinds, generating creative content, programming, translation, and the drafting of emails or essays (Javaid et al., 2023). In short, these tools can contribute significantly to increased productivity in both professional and academic settings (Chen et al., 2024). Their versatility and ease of use have led to a rapid and widespread adoption in universities, both among faculty and students. A study conducted with a sample of 800 university students in Spain found that 89% used AI applications for academic tasks, with consultation and conversational tools such as ChatGPT, Gemini, or DeepSeek being the most commonly used (81%) (Fundación CyD, 2025). Similarly, a study in Sweden reported that 95% of students were familiar with ChatGPT, and more than one-third used it regularly, while other chatbots were rarely used (Stöhr et al., 2024). Another study, focused on students in Saudi Arabia, found that 78.7% used GenAI tools for academic purposes (Almassaad et al., 2024).

Students consider chatbots -especially ChatGPT- to be useful, accessible, and motivating tools, particularly valued for their speed, interactivity, idea generation capabilities, immediate feedback, and ease of use (Schei et al., 2024). These AI tools offer a wide range of opportunities for both teaching and learning, including the design of tests and assessments, support for instructional materials, promotion of classroom collaboration, and their use as personalized tutors that adapt learning to each student's specific needs (Sok & Heng, 2024). They enable students and educators to access structured and written information quickly. Acting as efficient assistants, these tools can summarize relevant content, generate useful outputs, provide feedback aligned with the academic level and tone required, and support interactive dialogue that helps learners deepen their understanding, resolve doubts, and receive tailored corrections. This adaptability promotes meaningful, student-centered learning environments (Farrokhnia et al., 2024). Chatbots are also perceived as valuable for language learning, formulating research questions, and clarifying complex concepts (Monib et al., 2025). Moreover, they can function as writing assistants, translators, or style editors, relieving users of routine tasks and allowing them to focus on higher-order cognitive processes (Dwivedi et al., 2023). In sum, they are tools that enhance both productivity and creativity for students and instructors, and can foster autonomous learning when used effectively (Karafil & Uyar, 2025).

However, several concerns have also been raised. Indiscriminate use may limit students' critical thinking, creativity, and problem-solving abilities (Sok & Heng, 2024), particularly because these tools often return plausible-sounding answers that are not always accurate. Relying solely on models like ChatGPT to obtain factual and relevant information can hinder the development of analytical thinking and the transfer of knowledge to new contexts. Furthermore, the automation of academic tasks may weaken students' skills in interpretation, evaluation, and analysis, especially if they fail to verify or revise the generated content (Zirar, 2023). Consequently, errors, biases, or fabricated sources can compromise academic quality when outputs are not properly checked (Dwivedi et al., 2023). Ethical and legal concerns also arise, particularly regarding authorship, intellectual property, and accountability for AI-generated content. This highlights the need for regulatory frameworks that promote responsible and ethical use of such technologies. Accordingly, conceptual models have been proposed to encourage ethical chatbot use in educational settings (Chauncey & McKenna, 2023), as well as the reinforcement of good scholarly practices by authors and researchers (Cabezas-Clavijo et al., 2024; Kubota, 2023). Some authors have gone further, describing these tools as a threat -or even a plague- for education, due to their potential to erode core cognitive abilities and foster excessive dependence (Weissman, 2023). Concerns have also been raised about threats to academic integrity, particularly as chatbots may facilitate plagiarism and obscure the origin of AI-generated content (Sullivan et al., 2023).

Nevertheless, it is evident that AI is here to stay in educational settings, and the most reasonable approach is to integrate chatbots into learning experiences and processes in ways that support meaningful student learning. The meta-analysis by Wu & Yu (2024) indicates that AI chatbots have a significantly positive impact on multiple dimensions of learning, particularly among university students and in short-term interventions. However, their effectiveness depends on contextual factors, pedagogical design, and the reflective use of these tools by both students and educators. Among the recommendations for effective use of AI chatbots in education are fostering open dialogue about their role in learning environments and monitoring students' behavior in their interactions with AI agents (Jensen & Jensen, 2025). It is also advised to train users to operate these tools efficiently, and to review and revise the content they produce to ensure it is ethical and pedagogically appropriate (Karafil & Uyar, 2025). Chatbots should be understood as a means of promoting curiosity, autonomy, and engagement, but not as a substitute for intellectual effort or critical reflection. Through iterative use, these tools can help students construct knowledge from diverse perspectives and at varying levels of depth (Chauncey & McKenna, 2023). Therefore, it becomes necessary to revise curricular frameworks to adapt them to the new digital reality, adjusting syllabi, assignments, and assessment methods accordingly (Dwivedi et al., 2023; Farrokhnia et al., 2024).

***Generative Artificial Intelligence as a Tool for Bibliographic Reference Provision***

Since the launch of ChatGPT in November 2022, numerous studies have examined the ability of this and other subsequent chatbots to perform a range of academic tasks, including the generation of legitimate and relevant bibliographic references for research purposes. Several of these works have evaluated the reliability of a single version of ChatGPT in generating references. One study, analyzing a sample of 115 references in the biomedical field, found that 47% of the citations were entirely fabricated, while another 46% were real but contained significant errors in key bibliographic elements (Bhattacharyya et al., 2023). Similarly, Mugaanyi et al. (2024) assessed the capacity of ChatGPT-3.5 to generate valid references in both the natural sciences and the humanities. Based on a sample of 102 references, they found that approximately 75% corresponded to real sources, although only 32.7% of DOIs were correct in the sciences and just 8.5% in the humanities. The authors suggest that this may be due to the model's tendency to interpolate or "fill in" data when it cannot retrieve real information, rather than acknowledging the absence of such data. The study by Giray (2024) is also notable, as it analyzes the reliability of references generated by ChatGPT-3.5 depending on the type of document. Although it involved a small sample (10 articles and 10 books), all article references were fabricated, while all book references were real, although two of them included bibliographic inaccuracies.

Other studies have compared different versions of ChatGPT. For example, in a study focused on the field of library and information science, both ChatGPT-3.5 and ChatGPT-4o exhibited critical levels of hallucination, generating 96% and 94% false references, respectively (Oladokun et al., 2025). These models also displayed systematic errors, ranging from completely invented sources to incorrect details related to authorship, publication year, or journal title. Similarly, Walters and Wilder (2023) compared the accuracy of references generated by ChatGPT-3.5 and GPT-4 across 84 academic texts. Their results showed that 55% of the references from ChatGPT-3.5 were entirely fabricated, while GPT-4 reduced that proportion to 18%. However, issues persisted in specific types of sources; for instance, 70% of chapter citations generated by GPT-4 were false. The authors concluded that these deficiencies reflect a structural limitation in the way AI models generate references, underscoring the need for rigorous manual verification.

Finally, other studies have conducted comparative analyses of multiple chatbots, such as ChatGPT and Gemini. In the study by McGowan et al. (2023), which evaluated the ability of ChatGPT-3.5 and Bard (now Gemini) to generate valid bibliographic references in the field of psychiatry, ChatGPT produced only 6% correct references, while Bard generated none. Similarly, in a comparative analysis focused on generating references for systematic reviews in the medical field, Chelli et al. (2024) assessed the performance of ChatGPT-3.5, ChatGPT-4, and Bard (Gemini) in retrieving relevant and verifiable citations. Based on a sample of 471 references, the authors found extremely high rates of fabricated content, particularly in the case of Bard, which reached 91.4% fabricated citations. The different versions of ChatGPT also failed to demonstrate sufficient reliability to be used as primary sources in academic work. A later study by Omar et al. (2025) systematically evaluated GPT-4 and Gemini Ultra in the task of generating introductory sections for medical research articles, including credible and accurate references. Gemini yielded a higher proportion of correct references (77.2% versus 54% for ChatGPT); however, neither model proved fully reliable, as both produced fabricated or inaccurate citations. Notably, Gemini's hallucinations were more sophisticated, involving real author names and plausible structures, which made them harder to detect without manual verification. In another study, Aljamaan et al. (2024) proposed a scoring system to assess the authenticity of citations generated by AI chatbots in the medical domain. Comparing six AI tools (ChatGPT-3.5, Bard/Gemini, Perplexity, Bing -now Copilot-, Elicit, and SciSpace), the authors found that Gemini failed to generate a single valid citation, while ChatGPT-3.5 and Copilot showed high levels of fabrication. In contrast, academic-focused chatbots such as Elicit and SciSpace exhibited the lowest rates of fabricated references. Lastly, Spennemann (2025) compared four chatbots -ChatGPT-3.5, ChatGPT-4, ScholarGPT, and DeepSeek- and observed notable improvements in the more recent models, particularly ChatGPT-4 and DeepSeek, though none proved fully effective in generating entirely correct and legitimate citations. Moreover, the study highlighted a recurring issue: in many cases, the genuine sources cited across different chatbots originated from Wikipedia, revealing an excessive dependence on a single source. This overreliance may compromise both the quality and diversity of the information provided, especially considering that Wikipedia does not always meet rigorous academic standards.

Taken together, the literature review underscores that, despite the enthusiasm surrounding the use of AI chatbots in educational contexts, their ability to generate accurate and legitimate bibliographic references remains limited. Although newer models have shown improvements over earlier versions, structural issues persist, such as fabricated sources, metadata inaccuracies, and unreliable generation of unique identifiers. This highlights the need for empirical studies that systematically evaluate the quality and utility of references generated by various chatbots -not just ChatGPT- particularly in real-life use scenarios involving university students.

**Objectives**

The main objective of this study is to evaluate the reliability, accuracy, and relevance of bibliographic references provided by eight free-access artificial intelligence chatbots across five major areas of knowledge. The aim is to assess their suitability as academic support tools in higher education. This general objective is broken down into the following four specific objectives:

SO1. To identify the percentage of fully and partially correct references, as well as the proportion of fabricated references generated by the different AI models across the five major academic disciplines.

SO2. To analyze the characteristics of the references provided by the chatbots, including document type (book, article, others), average publication age, and the number of errors in bibliographic elements.

SO3. To measure the overlap between references generated by different AI tools and assess its potential impact on the diversity of academic sources available to students.

SO4. To identify the main works cited by the chatbots and evaluate their appropriateness for the specific academic task requested.

**Methods**

A comparative analysis was conducted to evaluate the performance of eight generative artificial intelligence chatbots- ChatGPT, Claude, Copilot, DeepSeek, Gemini, Grok, Le Chat, and Perplexity- in generating academic bibliographic references across the five major areas of knowledge (Health Sciences; Engineering/Technology; Experimental Sciences; Social Sciences; and Humanities). In all cases, the evaluation was carried out using the free-access versions of each chatbot, as most conventional users do not subscribe to the paid versions of these technologies. Table 1 presents the main features of the eight chatbot platforms included in the comparison.

**Table 1. Main characteristics of the chatbots compared**

| Chatbot | Company | Country | Launch date | Model |
|---|---|---|---|---|
| ChatGPT | Open AI | US | November-22 | GPT-4o-mini |
| Claude | Anthropic | US | March-23 | 3.5 Sonnet |
| Copilot | Microsoft | US | February-23 | GPT-4 |
| DeepSeek | DeepSeek | China | January-25 | DeepSeek-V3 |
| Gemini | Google | US | March-23 | Flash 2.0 |
| Grok | xAI | US | November-23 | Grok-3 |
| Le Chat | Mistral AI | France | February-25 | Mistral Large |
| Perplexity | Perplexity | US | December-22 | Sonar |

The tests were conducted between February 7 and 9, 2025. To simulate a realistic academic use scenario, a single standardized prompt was designed and adapted to five disciplines representative of each major area of knowledge: Cardiology (Health Sciences), Mechanical Engineering (Engineering/Technology), Organic Chemistry (Experimental Sciences), Sociology (Social Sciences), and Art History (Humanities). It is important to emphasize that all tests were performed using the free-access versions of each tool, as these are the versions most likely to be used by undergraduate and graduate students.

**Table 2. Prompts used by area of knowledge and discipline**

| Prompt | Broad Field | Subfield |
|---|---|---|
| "I am a university student working on my Final Degree Project. I need you to provide me with 10 relevant academic references in the field of Cardiology. Please format the references in APA 7th edition." | Health Sciences | Cardiology |
| "I am a university student working on my Final Degree Project. I need you to provide me with 10 relevant academic references in the field of Mechanical Engineering. Please format the references in APA 7th edition." | Engineering / Technology | Mechanical Engineering |
| "I am a university student working on my Final Degree Project. I need you to provide me with 10 relevant academic references in the field of Organic Chemistry. Please format the references in APA 7th edition." | Experimental Sciences | Organic Chemistry |
| "I am a university student working on my Final Degree Project. I need you to provide me with 10 relevant academic references in the field of Sociology. Please format the references in APA 7th edition." | Social Sciences | Sociology |
| "I am a university student working on my Final Degree Project. I need you to provide me with 10 relevant academic references in the field of Art History. Please format the references in APA 7th edition." | Humanities | Art History |

Each chatbot generated 50 references (10 per area of knowledge), resulting in a total sample of 400 references for analysis (80 per knowledge area). The variables analyzed were as follows:

- Reference accuracy (Completely Correct; Partially Correct; Wrong or Fabricated)
- Year of publication
- Document type (Article; Book; Other)
- Journal or publisher name
- Number of citations (according to Google Scholar as of March 27, 2025). In cases where a reference had multiple citation counts -such as a book with several editions- the highest value was recorded.

Additionally, to verify the accuracy of each reference, it was broken down into five core elements to determine which, if any, were incorrect:

- Author(s)
- Year
- Title of the work
- Publication venue (journal or publisher)
- Locating data (Volume, Issue, Pages, Digital Object Identifier – DOI)

Thus, each reference could contain between 0 and 5 errors. Verification was conducted through manual searches on Google and Google Scholar, enclosing the title provided by the chatbot in quotation marks.

Data processing and analysis were performed using Microsoft Excel, and graphical visualizations were created with SCImago Graphica 1.0.49 (Hassan-Montero et al., 2022).

**Results**

From the total dataset of 400 references analyzed, 26.5% were real and fully accurate (i.e., all five bibliographic elements were correct), while 33.8% were real but only partially correct (e.g., containing errors in the publication year or locating data). In contrast, 39.8% of the references were either incorrect or entirely fabricated by the AI systems. The chatbots that yielded the highest percentage of completely correct references were Grok (60%) and DeepSeek (48%). On the other hand, the chatbots most prone to fabricating references were Copilot (100%), Perplexity (72%), and Claude (64%). Notably, only two AIs -Grok and DeepSeek- did not fabricate any of the 50 references requested (Figure 1).

**Figure 1. Percentage of completely correct, partially correct, and incorrect or fabricated references, by AI chatbot**

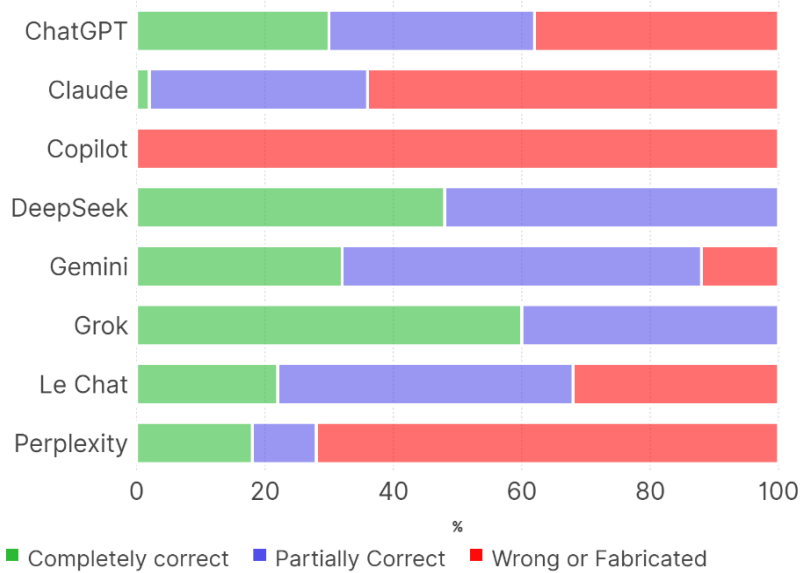

Each reference contained an average of two errors across the five parameters analyzed. Grok and DeepSeek recorded the lowest number of inaccurate elements, with 0.4 errors per reference for the former and 0.7 for the latter. In contrast, Copilot (4.2), Perplexity (3.3), and Claude (3.0) had the highest average number of missing or incorrect elements per reference (Figure 2).

**Figure 2. Distribution of errors per reference by AI chatbot**

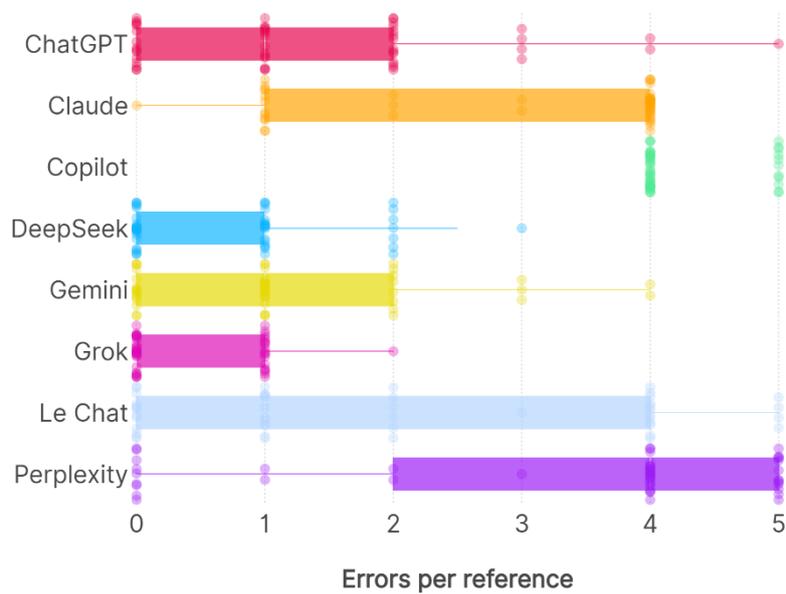

Regarding the document type of the generated references, 58.3% corresponded to books, 39.8% to journal articles, and 2% to other formats (such as book chapters, reports, or dictionary entries). Gemini was the AI that produced the highest proportion of book references (whether real or fabricated), at 90%, followed by DeepSeek (84%) and Grok (80%). In contrast, 100% of the references generated by Copilot were purported journal articles, although all were fabricated. Claude (66%) and Perplexity (54%) also showed a higher tendency to provide journal references, many of which were likewise fabricated (Figure 3).

**Figure 3. Percentage of references by document type and AI chatbot**

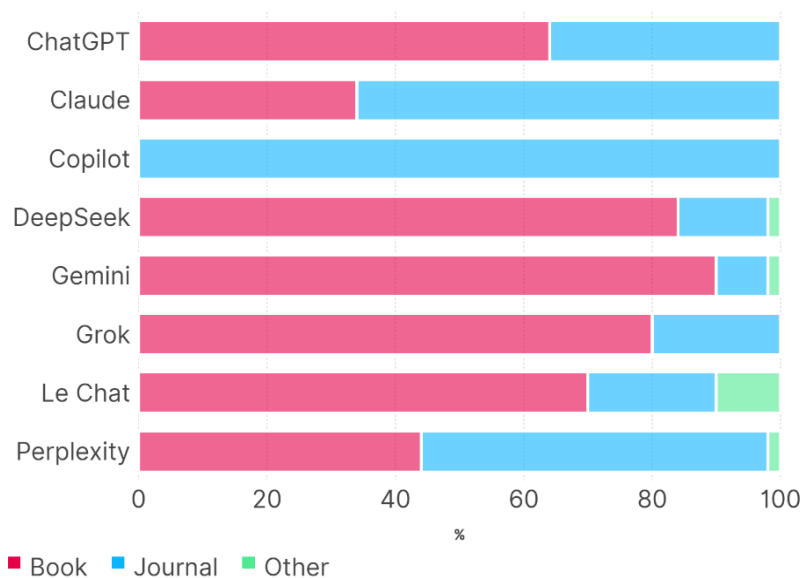

The highest percentage of fabricated references was found in the Engineering/Technology area, with 52.5%, followed by Health Sciences (50%) and Experimental Sciences (41.3%). These percentages were lower in Humanities (26.3%) and Social Sciences (28.8%) (Figure 4). These results are linked to the observation that the tendency to generate false references was significantly higher among journal articles than among books. While only 12.9% of book references were incorrect or fabricated, this figure rose to 78% in the case of journal references.

**Figure 4. Percentage of completely correct, partially correct, and wrong or fabricated references by area of knowledge**

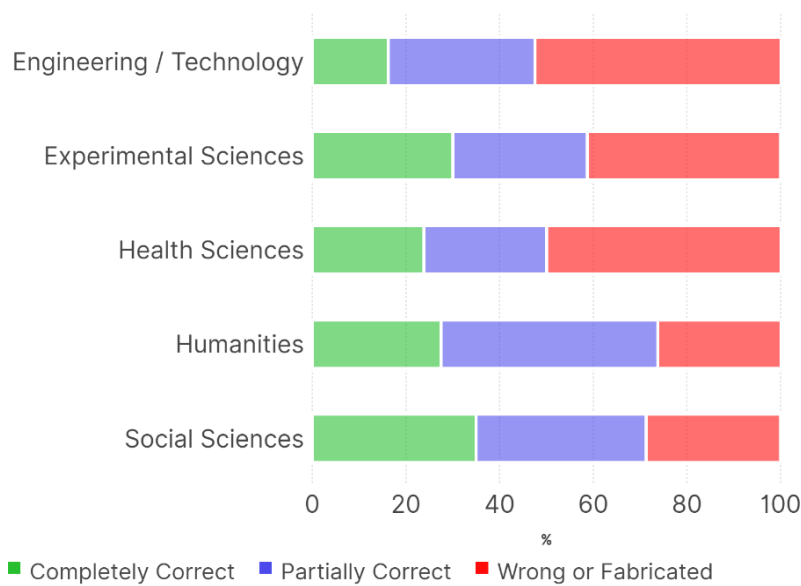

The average age of the references (including both real and fabricated) was 14.7 years. By far, the oldest references were generated by Le Chat, with an average of 45.3 years, including one reference from the 18th century and two from the 19th century. Gemini (25.9 years) and Grok (21.4 years) also presented high average ages. In contrast, Perplexity provided the most recent references, with an average age of 1.6 years, nearly three out of four of which were dated 2025, though the vast majority (72%) were fabricated. A similar pattern occurred with Claude, which generated very recent references (between 2019 and 2023) with an average age of 3.2 years, but with 64% of them being fabricated (Figure 5).

**Figure 5. Average publication age of references generated by each AI chatbot**

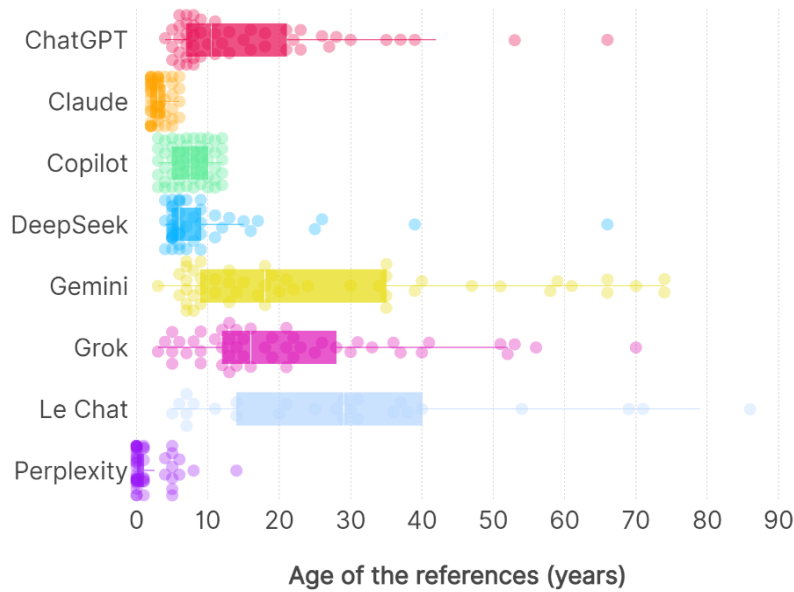

Regarding the average age of references by area of knowledge (including both real and fabricated sources), the oldest references were found in Social Sciences (average: 24.5 years) and Humanities (21.8 years), while the most recent references appeared in Engineering/Technology (7.4 years), Health Sciences (9.4 years), and Experimental Sciences (9.8 years) (Figure 6).

These figures are closely linked to the predominant document types within each area. The average age of journal references -which were more common in Engineering/Technology, Health Sciences, and Experimental Sciences- was 6.2 years across the entire sample. In contrast, book references, which were more frequent in Social Sciences and Humanities, had an average age of 20.3 years.

**Figure 6. Average publication age of references generated by area of knowledge**

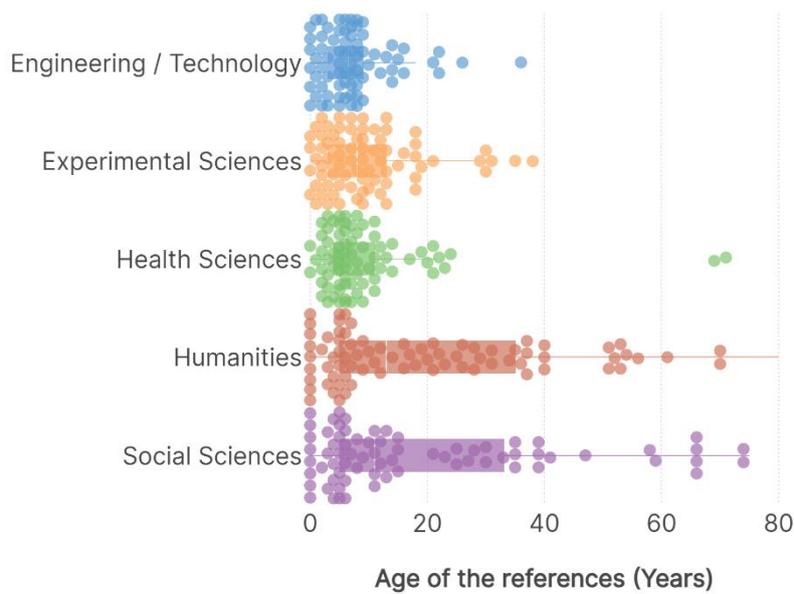

With respect to overlapping among AIs, there was a notably high degree of agreement among the real references provided by DeepSeek, Grok, Gemini, and ChatGPT. Specifically, 45% of the references generated by both DeepSeek and Grok, and 35% of those produced by Gemini, were also suggested by ChatGPT. Likewise, 36% of the references retrieved by DeepSeek were also provided by Gemini. Overall, the data reveal a high rate of overlap among these four chatbots, whereas the level of overlap with -and among- the remaining tools analyzed was significantly lower (Table 3).

**Table 3. Percentage of overlap among AI chatbots based on real references provided**

| | A/B | ChatGPT | Claude | Copilot | DeepSeek | Gemini | Grok | Le Chat | Perplexity |
|---|---|---|---|---|---|---|---|---|---|
| **B** | ChatGPT | | 3% | 0% | 45% | 35% | 45% | 3% | 0% |
| | Claude | 6% | | 0% | 11% | 0% | 0% | 17% | 0% |
| | Copilot | 0% | 0% | | 0% | 0% | 0% | 0% | 0% |
| | DeepSeek | 28% | 4% | 0% | | 32% | 32% | 6% | 0% |
| | Gemini | 25% | 0% | 0% | 36% | | 27% | 5% | 0% |
| | Grok | 28% | 0% | 0% | 32% | 24% | | 6% | 2% |
| | Le Chat | 3% | 9% | 0% | 9% | 6% | 9% | | 3% |
| | Perplexity | 0% | 0% | 0% | 0% | 0% | 7% | 7% | |

(Header row above data: spanning "A" across all chatbot columns.)

As for the sources of information provided by the AIs -whether real or fabricated- most originated from major academic publishers, such as Wiley (30 references), Oxford University Press (25 references), and McGraw Hill (19 references). The most frequently cited journal was Circulation (11 references), followed by Studies in the History of Art and the European Heart Journal, each appearing in 9 references. However, when fabricated references are excluded, Studies in the History of Art disappears from the list, and the number of real references linked to the leading publishers and journals declines slightly (Table 4).

**Table 4. Main sources provided by AI chatbots**

| ALL REFERENCES | | | REAL REFERENCES | | |
|---|---|---|---|---|---|
| Source | Type | N | Source | Type | N |
| Wiley | Publisher | 30 | Wiley | Publisher | 29 |
| Oxford University Press | Publisher | 25 | Oxford University Press | Publisher | 25 |
| McGraw-Hill | Publisher | 19 | McGraw-Hill | Publisher | 19 |
| Pearson | Publisher | 17 | Pearson | Publisher | 16 |
| Circulation | Journal | 11 | Routledge | Publisher | 10 |
| Routledge | Publisher | 10 | Elsevier | Publisher | 9 |
| Studies in the History of Art | Journal | 9 | Circulation | Journal | 9 |
| European Heart Journal | Journal | 9 | Princeton University Press | Publisher | 8 |
| Princeton University Press | Publisher | 8 | University of Chicago Press | Publisher | 8 |
| Elsevier | Publisher | 8 | Cengage Learning | Publisher | 7 |
| Journal of Organic Chemistry | Journal | 8 | European Heart Journal | Journal | 7 |
| Cengage Learning | Publisher | 7 | Journal of the American College of Cardiology | Journal | 6 |
| Journal of the American College of Cardiology | Journal | 7 | Phaidon Press | Publisher | 6 |

Table 5 presents the most frequently mentioned references, that is, those cited by at least three different AIs, along with the number of citations each has received. It is evident that the most commonly selected references by the chatbots also enjoy a very high number of citations in the academic literature -most with several thousand- indicating that these are intellectually relevant and authoritative sources. Moreover, with only one exception, all frequently selected references are monographs, many of which are manuals or textbooks. The most frequently selected work, identified by five different chatbots, is *March's Advanced Organic Chemistry: Reactions, Mechanisms, and Structure*, while six other references -all books- were chosen by four different AIs.

**Table 5. Most frequently cited bibliographic references across AI chatbots**

| Title[1] | Chatbots | Year | Journal or Publisher | Document type | Citations | Field |
|---|---|---|---|---|---|---|
| March's advanced organic chemistry: Reactions, mechanisms, and structure | 5 | 2007 | Wiley | Book | 16634 | Experimental Sciences |
| Organic chemistry (Clayden et al.) | 4 | 2012 | Oxford University Press | Book | 5068 | Experimental Sciences |
| Organic chemistry (McMurry) | 4 | 2016 | Cengage Learning | Book | nd | Experimental Sciences |

---

[1] In cases where the same book appeared in multiple editions, the year and title of the earliest edition were recorded. For the citation count, the highest number among the different versions of the same work was used.

| Title | | Year | Publisher | Type | Citations | Area |
|---|---|---|---|---|---|---|
| Shigley's mechanical engineering design | 4 | 2011 | McGraw-Hill | Book | 4729 | Experimental Sciences |
| The sociological imagination | 4 | 1959 | Oxford University Press | Book | 27784 | Social Sciences |
| The story of art (Gombrich) | 4 | 2002 | Phaidon Press | Book | 3466 | Humanities |
| Vision and Difference: Femininity, Feminism, and Histories of Art | 4 | 1988 | Routledge | Book | 3057 | Humanities |
| 2019 ESC guidelines for the diagnosis and management of chronic coronary syndromes | 3 | 2019 | European Heart Journal | Journal | 8421 | Health Sciences |
| Advanced organic chemistry: Part A: Structure and mechanism | 3 | 2007 | Springer | Book | 44 | Experimental Sciences |
| Braunwald's heart disease: A textbook of cardiovascular medicine | 3 | 2021 | Elsevier | Book | 8445 | Health Sciences |
| Distinction: A social critique of the judgement of taste | 3 | 1984 | Harvard University Press | Book | 88609 | Social Sciences |
| History of art: The Western tradition | 3 | 2004 | Prentice Hall | Book | 148 | Humanities |
| Materials science and engineering: An introduction | 3 | 2018 | Wiley | Book | 22930 | Engineering / Technology |
| Mechanics of materials | 3 | 2012 | McGraw-Hill | Book | 106 | Engineering / Technology |
| Organic chemistry (Bruyce) | 3 | 2014 | Pearson | Book | 1842 | Experimental Sciences |
| Studies in Iconology: Humanistic Themes in the Art of the Renaissance | 3 | 1939 | Oxford University Press | Book | 745 | Humanities |
| The division of labor in society | 3 | 1997 | Free Press | Book | 28699 | Social Sciences |
| The originality of the avant-garde and other modernist myths | 3 | 1985 | MIT Press | Book | 2749 | Humanities |
| The rise of the network society | 3 | 2010 | Wiley | Book | 57975 | Social Sciences |
| Ways of seeing | 3 | 1972 | Penguin Books | Book | 16755 | Humanities |

**Discussion and conclusions**

This study analyzed the performance of eight general-purpose chatbots in generating academic bibliographic references across five major areas of knowledge. All evaluations were conducted using the free-access versions of each tool, as this represents the most plausible use scenario for university students seeking assistance with bibliographic references for essays or academic assignments.

The overall results show that six out of the eight chatbots -all except Grok and DeepSeek- fabricated a substantial portion of the references requested (39.8%). This phenomenon of hallucination in reference generation has previously been documented for ChatGPT (Day, 2023; Giray, 2024; Walters & Wilder, 2023), but has rarely been analyzed in other chatbots. Compared to earlier studies, and specifically in relation to ChatGPT, the hallucination rate observed in our study represents a notable reduction from 55% in Walters & Wilder (2023), 57.7% in Spennemann (2025), and 100% in Day (2023) to 38% in our sample.

Our findings on DeepSeek, which did not fabricate a single reference, align with previous results, such as the 7% hallucination rate reported by Spennemann (2025). As for Gemini (formerly Bard), the data indicate a clear improvement, with the fabrication rate dropping from 100% in 2023 (McGowan et al., 2023) to 22.8% in 2024 (Omar et al., 2025), and now to 12% in our study. For other conversational chatbots such as Grok, Claude, and Le Chat, this is the first time performance data on academic reference generation has been reported.

From the perspective of academic rigor, it is concerning that only about one in four references generated was entirely correct- including authorship, year, title, publication venue, and locating data. This underscores that, while AI systems are becoming increasingly capable of generating legitimate bibliographic references, they are still far from being reliable tools for this fundamental academic task. Grok, with 60% fully correct references, and DeepSeek, with 48%, were the most trustworthy chatbots. This also aligns with the fact that neither of these two tools fabricated any references.

Fabrication levels also vary significantly across AI systems. While chatbots like ChatGPT and Gemini tend to combine real elements -such as author names or article titles- with invented data or information drawn from unrelated sources (resulting in plausible but false references), in other cases -such as Copilot- the references are entirely fabricated. For example, Microsoft's chatbot generated the same 10 references for all five knowledge areas, repeating author names, publication years, and locating information, and only changing the article titles and source names. As an illustrative example, Table 6 presents five fabricated references produced by this AI across different academic fields.

**Table 6. Examples of fabricated bibliographic references generated by Copilot**

| Field | Reference |
|---|---|
| Engineering / Technology | Green, M. L., & Brown, T. J. (2013). Advances in Mechanical Design. Journal of Mechanical Design, 27(4), 201-215. |
| Experimental Sciences | Green, M. L., & Brown, T. J. (2013). Advances in Organic Photovoltaics. Journal of Organic Photovoltaics, 27(4), 201-215. |
| Health Sciences | Green, M. L., & Brown, T. J. (2013). Peripheral Vascular Diseases: Current Concepts and Treatment. Journal of Vascular Medicine, 27(4), 201-215. |
| Humanities | Green, M. L., & Brown, T. J. (2013). French Genre Painting in the Eighteenth Century. Studies in the History of Art, 27(4), 201-215. |
| Social Sciences | Green, M. L., & Brown, T. J. (2013). Social Movements and Their Impact on Society. Journal of Social Movements, 27(4), 201-215. |

Although in the case of Copilot the fabrication pattern is relatively easy to detect at a glance, in other instances it is far more subtle, as the repetition is less obvious and fabricated references are even intermixed with real ones. A common mark for identifying false references is the frequent appearance of generic surnames (e.g., Smith, Lee, Brown, Wang, García) or overly generic journal and article titles. For this reason, it is essential to verify the accuracy of each reference, as recommended by Giray (2024). This verification is often as simple as performing a search for the exact title in quotation marks using either a general or academic search engine. Nonetheless, this verification step is frequently overlooked by students -and even by academics- as shown by the inclusion of fabricated references in some published scientific studies (Orduña-Malea & Cabezas-Clavijo, 2023).

Our study also found that the type of document is a key factor affecting the accuracy of the references. While 78% of journal article references were fabricated, this figure dropped to 12.9% in the case of book references. This pattern had already been observed by Giray (2024), who reported that ChatGPT generated 100% real book references (albeit with some bibliographic errors) and none for journal articles. Our findings reinforce this conclusion, and also reveal that chatbots are more likely to generate monograph references than scientific journal articles (58.3% books vs. 39.8% articles). This suggests that one of the key parameters guiding reference generation in chatbots is the salience or recurrence of a work in the model's training data. Although we measured the citation count and found it to be high for most real works retrieved, it is clear that citation metrics are not directly used in the generation process. However, it appears that the frequency with which a title appears in the training corpus -as is the case with highly cited works- is indeed a relevant factor in whether the AI retrieves or fabricates a given reference.

Another noteworthy aspect is that most of the books referenced by the chatbots have multiple published editions, some of which are available in full-text online. It should be noted that some of this content is accessible through non-academic websites or platforms of questionable legality. This suggests that certain AIs may be drawing on data sourced from websites that do not hold copyright licenses for the materials they disseminate. This issue is currently a subject of controversy and legal dispute between major content providers and large technology companies (Grynbaum & Mac, 2023). The possibility that AIs are trained on data from legally ambiguous sources may also help explain why they provide more book references than journal articles, although this remains a hypothesis that requires further research to be confirmed or refuted. In this regard, the fact that major publishers like Wiley are already entering into agreements with technology companies to license their content (Battersby, 2024) is especially relevant. Such partnerships should lead to improved access to high-quality content via chatbots, and to greater accuracy in academic tasks such as the generation of real and verifiable bibliographic references. In return, publishers may benefit from increased web traffic driven by AI tools to their platforms. This dynamic may prove especially significant not only for large scientific publishers but also for smaller academic and non-profit presses, which often lack the resources to promote their content widely. Therefore, the findability and accessibility of a publisher's content through generative AI tools may become a critical factor in its use and citation, and should be considered a strategic priority in the evolving academic publishing landscape.

In the cases where the AI was able to identify or provide real references, many of them were incomplete or contained inaccurate information (33.8% of entries). One of the most frequent errors -in the case of books- was the mixing of data from different editions. As previously noted, most of the book references provided were relatively old (with an average age of 20.3 years) and corresponded to foundational works within each discipline, often in the form of textbooks or introductory manuals. These works tend to be reissued regularly, either to update their content or to add new material, sometimes under different publishers or with modified authorship. In such cases, the AI frequently blended information from various editions, such as publication years or author lists; thus, while the work itself was real, the reference was not accurate. This may lead to the systematic reproduction of referencing errors when using chatbots for citation tasks.

Additionally, we compared results across the five major academic areas to detect differences in reference generation. Two clear patterns emerged: first, in Experimental Sciences, Engineering and Technology, and Health Sciences -fields that typically rely more on journal articles than books- AI systems did indeed generate a greater number of article references. In contrast, in Social Sciences and Humanities, where monographs are more commonly used, chatbots provided an overwhelming majority of book references. This distinction is important, as the chatbots analyzed tend to hallucinate more journal article references than book references. Therefore, the information provided by chatbots in Social Sciences and Humanities appears to be more reliable than that offered in the other domains.

This phenomenon is also closely linked to the age of the references. In Social Sciences and Humanities, the average age of the references is significantly higher than in other academic areas. This is understandable, as knowledge in these fields tends to retain its validity over time and does not become obsolete as quickly as in disciplines with faster research cycles. However, we also observed a tendency among certain AIs -notably Le Chat, Gemini, and Grok- to provide very old references even in experimental fields. In such cases, while the references may be real, there may be issues regarding their relevance, as the content may be outdated or surpassed by newer techniques and findings. Moreover, several chatbots returned references that are unlikely to be read or used by undergraduate students, such as works published before the 20th century or texts with an excessively high level of specialization or complexity. In fact, the use of references that are too old or disproportionately advanced in student papers may be an indicator of uncritical reliance on AI-generated citations.

Another striking finding is the high degree of overlap among certain AIs in response to bibliographic queries. Considering that, as of 2019, there were an estimated 389 million bibliographic references accessible via Google Scholar (Gusenbauer, 2019), the level of coincidence in reference suggestions across ChatGPT, DeepSeek, Grok, and Gemini is remarkable. This suggests that these chatbots may share similar training data or reasoning models, which could explain the convergence in their outputs. This raises a broader concern: that AI-generated responses may be constrained to a narrow intellectual framework, largely reproducing established and widely cited knowledge, while limiting the diversity of perspectives presented. In doing so, generative AI may unintentionally reinforce existing paradigms and curb the capacity for innovation and critical thinking in society at large.

From an educational perspective, these results highlight the need to strengthen AI literacy policies within universities, particularly in relation to search techniques and critical evaluation of information. A majority of university students report using AI tools for class assignments, essays, final degree projects, and other academic tasks (Almassaad et al., 2024; Fundación CyD, 2025; Stöhr et al., 2024). Therefore, it is essential to enhance students' information literacy skills so they can make the most of AI applications, distinguishing between plausible yet incorrect information and content that is factual and aligned with their academic needs. The widespread use of AI tools across all sectors of society -and particularly in academia- calls for stronger institutional policies on information literacy, aimed at promoting ethical, efficient, and accurate use of the information generated by these technologies (Bhullar et al., 2024; Giray, 2024; Jensen & Jensen, 2025). Key topics such as information bias, the unchecked spread of errors, and the distribution of false or misleading content must form the foundation of a new AI-centered literacy framework, which builds upon the core competencies in information use expected of any university student. Our findings point to the importance of incorporating training on AI-generated bibliographic references into university curricula, in line with the recommendations proposed by Giray (2024). Ultimately, students must be taught how to verify information using instructional strategies that emphasize critical thinking (Bhullar et al., 2024). As noted earlier, the recurrent appearance of overly common surnames or excessively generic titles may serve as clear indicators that the references in question are fabricated and therefore unreliable.

It is also important to consider that this experimental study was conducted in a realistic scenario for university students, but not the only one. There are AI tools specifically designed for academic research and the discovery of validated references, such as Elicit or ResearchRabbit. However, these tools are unlikely to be widely known or effectively used by most students (Jensen & Jensen, 2025). Similarly, premium or paid versions of the various chatbots are not affordable for all students, particularly those in lower-income regions. This points to the existence of a potential "AI divide" (Daepp & Counts, 2025) between individuals who can afford access to more advanced chatbot versions -thus benefiting from enhanced capabilities- and those who rely solely on the free versions, which tend to be more limited and less reliable.

Finally, certain limitations of this study must be acknowledged. First, the sample size -400 references- is necessarily small, though sufficient to draw meaningful conclusions about the performance of the chatbots in bibliographic reference generation. Moreover, the rapid pace at which large language models (LLMs) are evolving, and the speed at which they are being deployed by major technology companies, mean that some of these findings may quickly become outdated. Nevertheless, it is worth noting that, despite the time elapsed since the launch of ChatGPT (November 2022), a high percentage of fabricated references still persists across many of the free-access AI chatbots tested. Future research could expand this analysis to more specific disciplinary contexts, include premium versions of the models, or explore specialized academic discovery tools, with the aim of designing safer, more ethical, and more effective learning ecosystems for university students.

In summary, this study provides relevant empirical evidence on the capabilities and limitations of artificial intelligence chatbots in the generation of academic bibliographic references- an essential task in the context of higher education. While certain tools such as Grok and DeepSeek demonstrate promising performance, the high rate of fabricated references found in most models highlights the risks associated with the uncritical use of AI-generated information by students. These findings underscore the need to strengthen pedagogical strategies focused on information literacy, particularly within the evolving educational framework shaped by artificial intelligence.

**GEN-AI USE**

ChatGPT 4.0 was used to improve the writing style and clarity of the text, as well as to produce the English translation of the manuscript from the original version written in Spanish.